\documentclass{elsart}

\usepackage{epsfig}
\usepackage{caption2}
\usepackage{amsmath}
\usepackage{graphicx}
\usepackage{graphics}

\newcommand{\ignore}[1]{}
\def\square{\vrule height6pt width7pt depth1pt}
\def\endpf{\hfill\square\bigskip}

\begin{document}

\begin{frontmatter}

% Title, authors and addresses

% use the thanksref command within \title, \author or \address for footnotes;
% use the corauthref command within \author for corresponding author footnotes;
% use the ead command for the email address,
% and the form \ead[url] for the home page:
% \title{Title\thanksref{label1}}
% \thanks[label1]{}
% \author{Name\corauthref{cor1}\thanksref{label2}}
% \ead{email address}
% \ead[url]{home page}
% \thanks[label2]{}
% \corauth[cor1]{}
% \address{Address\thanksref{label3}}
% \thanks[label3]{}

\title{Searching Monotone Multi-dimensional Arrays}

% use optional labels to link authors explicitly to addresses:
% \author[label1,label2]{}
% \address[label1]{}
% \address[label2]{}

\author[cyx]{Yongxi Cheng}
\author[sxm]{Xiaoming Sun}
\author[yyq]{Yiqun Lisa Yin}

\

\

\address[cyx]{Department of Computer Science, Tsinghua University, Beijing 100084, China \\ Email:
cyx@mails.tsinghua.edu.cn}
\address[sxm]{Center for Advanced Study, Tsinghua University,
Beijing 100084, China \\ Email: xiaomings@tsinghua.edu.cn}
\address[yyq]{Independent security consultant,
Greenwich CT, USA\\ Email: yiqun@alum.mit.edu}

\newpage

\begin{abstract}
% Text of abstract
In this paper we investigate the problem of searching monotone multi-dimensional arrays. We generalize Linial and
Saks' search algorithm~\cite{LS1} for monotone 3-dimensional arrays to $d$-dimensions with $d\geq 4$. Our new
search algorithm is asymptotically optimal for $d=4$.
\end{abstract}

\begin{keyword}
% keywords here, in the form: keyword \sep keyword
search algorithm \sep complexity \sep multi-dimensional array \sep partial order
% PACS codes here, in the form: \PACS code \sep code
%\PACS
\end{keyword}
\end{frontmatter}

\newtheorem{coro}{Corollary}[section]
\newtheorem{defi}{Definition}[section]
\newtheorem{exam}{Example}[section]
\newtheorem{lemm}{Lemma}[section]
\newtheorem{theo}{Theorem}[section]

% main text
\section{Introduction}

In this paper, we investigate the problem of searching monotone multi-dimensional arrays. Suppose we are given a
$d$-dimensional array with $n$ entries along each dimension
\[ A_{n,d}=\{a_{i_1,i_2,\ldots,i_d}|i_1,i_2,\ldots,i_d=1,2,\ldots,n\}.\]
We say that the array $A_{n,d}$ is {\em monotone} if its entries are real numbers that are increasing along each
dimension. More precisely, if $i_1\leq j_1$, $i_2\leq j_2$,\ldots, $i_d\leq j_d$ then $a_{i_1,i_2,\ldots,i_d}\leq
a_{j_1,j_2,\ldots,j_d}$. In other words, if $P=[n]^d$ is the product of $d$ totally ordered sets
$\{1,2,\ldots,n\}$, than $A_{n,d}$ is consistent with the partial order $P$.

The {\em search problem} is to decide whether a given real number $x$ belongs to the array $A_{n,d}$ by comparing
$x$ with a subset of the entries in the array. The {\em complexity} of this problem, denoted by $\tau(n,d)$, is
defined to be the minimum over all search algorithms for $A_{n,d}$ of the number of comparisons needed in the
worst case. Note that for $d=1$, this problem reduces to searching a totally ordered set. In this case, the binary
search algorithm is optimal and requires at most $\lceil \log_2 (n+1) \rceil$ comparisons in the worst case.

We first briefly review some previous work. In~\cite{LS2}, Linial and Saks presented some general results on the
complexity of the above class of search problems, for any finite partially ordered set $P$. In~\cite{LS1} they
studied the problems for general finite partially ordered set $P$ and also gave more precise results for the case
where $P=[n]^d$, for dimensions $d\geq 2$. They observed that for $d=2$, it had been known that
$\tau(n,2)=2n-1$~\cite{GK}. For the generalized case $d\geq 2$, they showed that the order of $\tau(n,d)$ is
$O(n^{d-1})$. More specifically, they proved that for $d\geq 2,$
\[ c_1(d)n^{d-1}\geq \tau(n,d)\geq c_2(d)n^{d-1}+o(n^{d-1}),\]
where $c_1(d)$ is a nonincreasing function of $d$ and upper bounded by 2,
and $c_2(d)=\sqrt{(24/\pi)}d^{-1/2}+o(d^{-1/2}).$ The upper bound
 $c_1(d)n^{d-1}$ was obtained by using a straightforward search
algorithm which partitions $A_{n,d}$ into $n$ isomorphic copies of
$A_{n,d-1}$ and searches each copy separately. They also described
a more efficient algorithm for $d=3$ and proved the following
bounds on $\tau(n,3)$:
\[ \lfloor\frac{3n^2}{2}\rfloor \le \tau(n,3) \le \frac{3n^2}{2} + cn \ln n. \] In the above inequality, $c$ is a positive constant,
and so the bounds are asymptotically tight. An open problem left is whether their search algorithm for $d=3$ can
be generalized to higher dimensions.

In this paper, we present new search algorithms for monotone $d$-dimensional arrays for $d \ge 4$, by generalizing
the techniques in~\cite{LS1} to higher dimensions. For $d\geq4$, the search complexity of our algorithms is
\[ \tau(n,d) \le \frac{d}{d-1} n^{d-1}+O(n^{d-2}).\]
The above bound is tight for $d=4$, up to the lower order terms.

The rest of the paper is devoted to the description and analysis of
the new algorithms. We start with the case where $d=4$. This special
case best illustrates the main idea, and it is also easier to visualize the subspaces that are encountered in the search algorithm. Then we describe the generalized  algorithm for $d\geq 4$.

Before presenting the technical details, we describe some basic notation and convention that we will follow
throughout the paper. In general, we use capital letters to represent sets and small letters to represent numbers.
The sets that we need to consider are often subsets of $A_{n,d}$ for which some of the subscripts are fixed, and
we use some simple notation to represent them. For example, we use $Q=\{a_{1,i_2,i_3,i_4}\}$ to denote a
``surface'' of the $4$-dimensional array $A_{n,4}$ for which the first subscript of $a$ is fixed to be 1. It is
understood that all other subscripts range between $[1,n]$, and we often omit the specification
``$i_2,i_3,i_4=1,2,\ldots,n$" if it is clear from the context.

\section{Searching $4$-Dimensional Arrays}

In this section, we present a $\frac{4}{3}n^3+O(n^2)$ algorithm
for searching monotone $4$-dimensional arrays. The algorithm is optimal
up to the lower order terms.

We start with a lower bound on $\tau(n,4)$ which will be seen asymptotically tight later, followed by the
description of an algorithm for partitioning monotone two-dimensional arrays, which will be a useful subroutine
for our searching algorithm. Then, we will present the main idea and the details of our search algorithm for
4-dimensional arrays.

\subsection{A lower bound on $\tau(n,4)$}

Using the method in \cite{LS1}, we can calculate a lower bound on
$\tau(n,4)$. Let $[n]$ denote the totally ordered set
$\{1,2,\ldots,n\}$, and let
\begin{equation*}
\begin{aligned}
D_1(n,4) & = & \{(i_1,i_2,i_3,i_4)\in [n]^4|i_1+i_2+i_3+i_4=2n+1\},\\
D_2(n,4) & = & \{(i_1,i_2,i_3,i_4)\in [n]^4|i_1+i_2+i_3+i_4=2n+2\}.
\end{aligned}
\end{equation*}
Define $D(n,4)=D_1(n,4)\cup D_2(n,4)$. Then $D(n,4)$ is a
\textit{section} (see~\cite{LS1}) of $[n]^4$, and there is no
ordered chain having length more than 2 in $D(n,4)$. Therefore,
$\tau(n,4)$ is lower bounded by $|D(n,4)|$. Let
\begin{eqnarray*}
X &=& \{(i_1,i_2,i_3,i_4)\in [2n+1]^4| \,\, i_1+i_2+i_3+i_4=2n+1\},\\
Y_k &=& \{(i_1,i_2,i_3,i_4)\in X| \,\, i_k>n\} \mbox{  for $k=1,2,3,4$},\\
Z &=& \{(i_1,i_2,i_3,i_4)\in [n+1]^4| \,\, i_1+i_2+i_3+i_4=n+1\}.
\end{eqnarray*}
It is easy to see that $|Y_k|=|Z|={n
\choose 3}$ for $k=1,2,3,4$. Thus,
$|D_1(n,4)|=|X|-\sum_{k=1}^{4}|Y_k|={2n \choose 3}-4{n \choose
3}=\frac {1}{3}(2n^3-2n)$. Similarly, $|D_2(n,4)|={2n+1 \choose
3}-4{n+1 \choose 3}=\frac {1}{3}(2n^3+n)$. Therefore,
\[\tau(n,4) \ge |D(n,4)|=|D_1(n,4)|+|D_2(n,4)|=\frac{4}{3}n^3-\frac{n}{3}.\]

\subsection{Partitioning 2-dimensional arrays}
\label{par}

In~\cite{LS1}, Linial and Saks gave a simple search algorithm for an $m\times n$ matrix ($m,n\geq 1$) with entries
increasing along each row and column. The algorithm needs at most $m+n-1$ comparisons. We will refer to this
algorithm as the {\it Matrix Search Algorithm}. Since $A_{n,2}$ is isomorphic to an $n\times n$ matrix, we can
adapt the {\it Matrix Search Algorithm} to partition $A_{n,2}$ into two subsets $S$ and $L$ given an input $x$,
such that $S$ contains entries \emph{smaller} than $x$ and $L$ contains entries \emph{larger} than $x$, using at
most $2n-1$ comparisons. Below, we provide the detailed description of the partition algorithm for the sake of
completeness.

\noindent \textit{\textbf{Algorithm: }Partition 2-Dimensional Array}

\noindent \textit{Input.}
\begin{itemize}
\item A real number $x$.
\item A monotone 2-dimensional array $A_{n,2}=\{a_{i_1,i_2}\}$.
\end{itemize}
\noindent \textit{Output.}
\begin{itemize}
\item If $x\in A_{n,2}$, output $(i_1,i_2)$ such that $a_{i_1,i_2}=x$.
\item If $x\not\in A_{n,2}$, output a partition $\{u,v,S,L\}$ of $A_{n,2}$
with the following properties:\\
-- $u$ and $v$ are two arrays each contains $n$ integers such that
$i_1\leq u[i_2]$ iff $a_{i_1,i_2}<x$ and $i_2\leq v[i_1]$ iff $a_{i_1,i_2}<x$.\\
-- $S$ and $L$ form a partition of $\{(i_1,i_2)| i_1,i_2 \in [n]\}$ such that if $(i_1,i_2) \in S$ then
$a_{i_1,i_2}<x$, and if $(i_1,i_2) \in L$ then $a_{i_1,i_2}>x$.
\end{itemize}

\noindent \textit{Procedure.}
\begin{itemize}
\item Initially set $S=L=\phi$.
\item View  $A_{n,2}$ as an $n\times n$ matrix and repeat
comparing $x$ with the element $e$ at the top right corner of the current matrix.\\
--  If $x>e$, then eliminate the first row of the current matrix and put their entries into $S$;\\
--  If $x<e$, then eliminate the last column of the current matrix and put their entries into $L$;\\
--  If $x=e$, then return this entry and exit.
\item Stop when the partition is finished, thus also obtain $u$ and $v$ (see Fig. \ref{fig:2d}).
\end{itemize}

We will use the notation $u$,$v$,$S$,$L$ throughout the paper.
Sometimes we will introduce subscripts to them to represent the
dimension indices to be considered. Ignoring the indices, these four
variables have the following useful relations:

\begin{figure}[t]
\renewcommand{\captionlabelfont}{\bf}
\renewcommand{\captionlabeldelim}{.~}
\centering
\includegraphics[width=100mm]{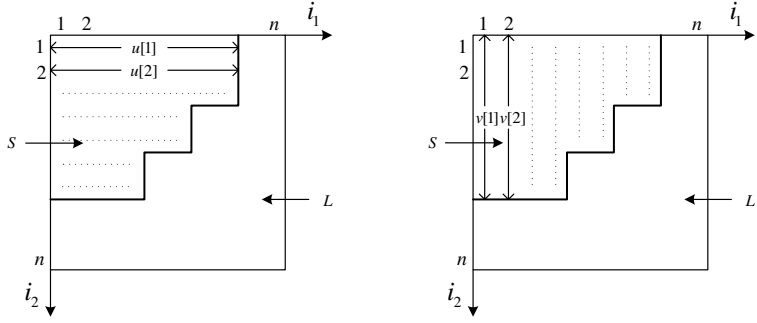}
\renewcommand{\figurename}{\textbf{Fig.}}
\caption{$u,v,S,L$: partition of the monotone 2-dimensional array $A_{n,2}$} \label{fig:2d}
\end{figure}

\begin{equation}
\label{equ:Suv} S = \{(i_1,i_2)|1\leq i_1\leq
u[i_2]\}=\{(i_1,i_2)|1\leq i_2\leq v[i_1]\}
\end{equation}

\begin{equation}
\label{equ:Luv} L = \{(i_1,i_2)|u[i_2]<i_1\leq n\}=\{(i_1,i_2)|
v[i_1]<i_2\leq n\}.
\end{equation}

Obviously, $S\cap L=\phi$ and $S\cup L=[n]^2$, hence $|S|+|L|=n^2$.
In addition, $|S|=u[1]+\ldots+u[n]=v[1]+\ldots+v[n]$ and
$|L|=(n-u[1])+\ldots+(n-u[n])=(n-v[1])+\ldots+(n-v[n])$.

Notice that when $m=0$ or $n=0$, we can ``search'' an $m\times n$ matrix using $0$ comparisons. Therefore, based
on the {\it Matrix Search Algorithm} in~\cite{LS1}, we have the following lemma that will be useful later.
\begin{lemm}
\label{searchmatrix} For $m,n\geq 0$, any $m\times n$ matrix with
entries increasing along each dimension can be searched using at
most $m+n$ comparisons.
\end{lemm}
{\bf Proof.}\, If $m=0$ or $n=0$, the matrix is empty, thus needs no comparison. If $m,n>0$, using the {\it Matrix
Search Algorithm}, we can search the matrix using at most $m+n-1$ comparisons. Therefore, the lemma holds. \endpf

\subsection{Main idea of the search algorithm}

\label{mainidea}

The main idea of our algorithm for $d=4$ is to first search the
``surfaces" ($3$-dimensional arrays) of $A_{n,4}$ and then the
problem reduces to searching a ``smaller" array $A_{n-2,4}$. At a
high level, searching the surfaces consists of two major steps:
\begin{itemize}
\item Step 1: Select $8$ special $2$-dimensional arrays and
partition each into two subsets $L$ and $S$, where elements in $L$
are larger than or equal to $x$, and elements in $S$ are smaller
than $x$, using the algorithm \emph{Partition 2-Dimensional Array.}
\item Step 2: Search the $8$ ``surfaces" of $A_{n,4}$. The subsets
$S, L$ obtained in Step 1 help to ``cut" each surface into a
sequence of 2-dimensional matrices that allows searching with less
comparisons.
\end{itemize}

\subsection{Description and analysis of the search algorithm}

Now we are ready to present our search algorithm for $d=4$.
As explained in Section~\ref{mainidea}, the algorithm is recursive,
which reduces $n$ by two for each recursion.
Without loss of generality, we consider the case where $x\not \in
A_{n,4}$. We first describe the algorithm and then analyze the number
of comparisons needed.

\noindent \emph{Step 1.} \ Apply the algorithm \textit{Partitioning
2-Dimensional Array} to divide each of the following eight
2-dimensional arrays into two subsets (the eight arrays are defined
by fixing two of the subscripts to either 1 or $n$, thus reducing
the number of dimensions by two).

\begin{equation*}
\begin{aligned}
M_1 & = & \{a_{i_1,i_2,1,n}\}:\,\,\,S_1, L_1;\ \ \ \ \ \ \ \ \ \ \ M^{*}_1=\{a_{i_1,i_2,n,1}\}:\,\,\,S^{*}_1, L^{*}_1;\\
M_2 & = & \{a_{n,i_2,i_3,1}\}:\,\,\,S_2, L_2;\ \ \ \ \ \ \ \ \ \ \ M^{*}_2=\{a_{1,i_2,i_3,n}\}:\,\,\,S^{*}_2, L^{*}_2;\\
M_3 & = & \{a_{1,n,i_3,i_4}\}:\,\,\,S_3, L_3;\ \ \ \ \ \ \ \ \ \ \ M^{*}_3=\{a_{n,1,i_3,i_4}\}:\,\,\,S^{*}_3, L^{*}_3;\\
M_4 & = & \{a_{i_1,1,n,i_4}\}:\,\,\,S_4, L_4;\ \ \ \ \ \ \ \ \ \ \
M^{*}_4=\{a_{i_1,n,1,i_4}\}:\,\,\,S^{*}_4, L^{*}_4.
\end{aligned}
\end{equation*}
The eight pairs of ``mutually complementary'' subsets $S_k, L_k$ and
${S_k}^{*}, {L_k}^{*}$ ($k=1,2,3,4$) have the following properties:
\begin{equation*}
\begin{aligned}
a_{i_1,i_2,1,n}<x<a_{j_1,j_2,1,n} & & \mbox{\ \ \ \ \ \ for $(i_1,i_2)\in S_1$ and  $(j_1,j_2)\in L_1$}\\
\\
a_{n,i_2,i_3,1}<x<a_{n,j_2,j_3,1} & & \mbox{\ \ \ \ \ \ for $(i_2,i_3)\in S_2$ and  $(j_2,j_3)\in L_2$}\\
\\
a_{1,n,i_3,i_4}<x<a_{1,n,j_3,j_4} & & \mbox{\ \ \ \ \ \ for $(i_3,i_4)\in S_3$ and $ (j_3,j_4)\in L_3$}\\
\\
a_{i_1,1,n,i_4}<x<a_{j_1,1,n,j_4} & & \mbox{\ \ \ \ \ \ for $(i_4,i_1)\in S_4$
and  $(j_4,j_1)\in L_4$}
\end{aligned}
\end{equation*}
%half

\begin{equation*}
\begin{aligned}
a_{i_1,i_2,n,1}<x<a_{j_1,j_2,n,1} & & \mbox{\ \ \ \ \ \ for $(i_1,i_2)\in S^{*}_1$ and $(j_1,j_2)\in L^{*}_1$}\\
\\
a_{1,i_2,i_3,n}<x<a_{1,j_2,j_3,n} & & \mbox{\ \ \ \ \ \ for $(i_2,i_3)\in S^{*}_2$ and  $(j_2,j_3)\in L^{*}_2$}\\
\\
a_{n,1,i_3,i_4}<x<a_{n,1,j_3,j_4} & & \mbox{\ \ \ \ \ \ for $(i_3,i_4)\in S^{*}_3$ and $(j_3,j_4)\in L^{*}_3$}\\
\\
a_{i_1,n,1,i_4}<x<a_{j_1,n,1,j_4} & & \mbox{\ \ \ \ \ \ for $(i_4,i_1)\in
S^{*}_4$ and $(j_4,j_1)\in L^{*}_4$}
\end{aligned}
\end{equation*}

In addition to the eight pairs of subsets, the algorithm also
outputs $u_k, v_k$ and $u^*_k,v^*_k$, corresponding to $S_k, L_k$
and ${S_k}^{*}, {L_k}^{*}$ respectively, with the properties given
in Equation~\ref{equ:Suv} and~\ref{equ:Luv}. For each $k$, at most
$2n-1$ comparisons are needed to partition $M_k$ ($M^{*}_k$). Thus,
at most $8 \times (2n-1)$ comparisons are needed in this step.

\noindent \emph{Step 2.} \ Search the following eight 3-dimensional
surfaces of $A_{n,4}$ (each surface is defined by setting one of the
subscripts to either $1$ or $n$, thus reducing the number of
dimensions by one).

\begin{equation*}
\begin{aligned}
Q_1 & = & \{a_{1,i_2,i_3,i_4}\}\ \ \ \ \ \ \ \ \ \ \ \ \ \ {Q^{*}_1}=\{a_{n,i_2,i_3,i_4}\}\\
Q_2 & = & \{a_{i_1,1,i_3,i_4}\}\ \ \ \ \ \ \ \ \ \ \ \ \ \ {Q^{*}_2}=\{a_{i_1,n,i_3,i_4}\}\\
Q_3 & = & \{a_{i_1,i_2,1,i_4}\}\ \ \ \ \ \ \ \ \ \ \ \ \ \ {Q^{*}_3}=\{a_{i_1,i_2,n,i_4}\}\\
Q_4 & = & \{a_{i_1,i_2,i_3,1}\}\ \ \ \ \ \ \ \ \ \ \ \ \ \ {Q^{*}_4}=\{a_{i_1,i_2,i_3,n}\}
\end{aligned}
\end{equation*}

By symmetry, we only need to show how to search $Q_1$. The algorithm
proceeds by fixing $i_3=i_3^{'}$ for $i_3^{'}=1,2,...,n$ and
searching each of the 2-dimensional array
$\{a_{1,i_2,i_3^{'},i_4}\}$. A useful observation is that for each
$i_3^{'}$, we can restrict the search to a smaller matrix (in
contrast to an $n\times n$ matrix) by leveraging on information
obtained in Step~1.

\begin{figure}[t]
\renewcommand{\captionlabelfont}{\bf}
\renewcommand{\captionlabeldelim}{.~}
\centering
\includegraphics[width=135mm]{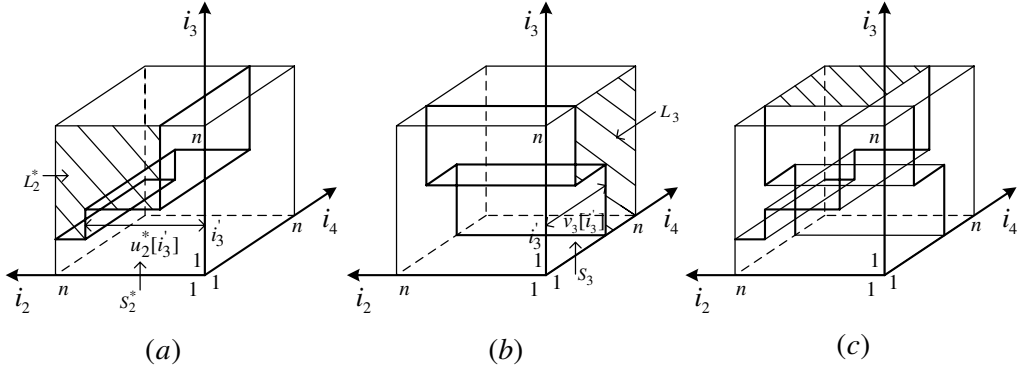}
\renewcommand{\figurename}{\textbf{Fig.}}
\caption{Searching the 3-dimensional surface
$Q_1=\{a_{1,i_2,i_3,i_4}\}$ of $A_{n,4}$. (a): partition of
$M^{*}_2=\{a_{1,i_2,i_3,n}\}$ into $S^{*}_2$ and $L^{*}_2$; (b):
partition of $M_3= \{a_{1,n,i_3,i_4}\}$ into $S_3$ and $L_3$; (c):
the `pyramid' composed of a sequence of 2-dimensional matrices to be
searched.} \label{fig:3d}
\end{figure}

Below, we explain the above observation and Step~2 in more details.
Consider an element $a_{1,i_2,i_3^{'},i_4}\in Q_1$. If
$(i_2,i_3^{'})\in S^{*}_2$, then we know that
$a_{1,i_2,i_3^{'},i_4}\leq a_{1,i_2,i_3^{'},n} <x.$ Hence, in order
for $a_{1,i_2,i_3^{'},i_4}=x$, it must be the case that
$(i_2,i_3^{'})\in L^{*}_2$, or equivalently,
$u^{*}_2[i_3^{'}]<i_2\leq n$. Similarly, we can conclude that in
order for $a_{1,i_2,i_3^{'},i_4}=x$, it must be the case that
$(i_3^{'},i_4)\in L_3$, or equivalently, $v_3[i_3^{'}]<i_4\leq n$.
Hence, we obtain a constraint on the indices $(i_2,i_4)$. By
lemma~\ref{searchmatrix}, searching this  restricted
$(n-u^{*}_2[i_3^{'}])\times (n-v_3[i_3^{'}])$ matrix needs at most
$(n-u^{*}_2[i_3^{'}])+(n-v_3[i_3^{'}])$ comparisons. Notice that
$n-u^{*}_2[i_3^{'}]$ is the number of entries $(i_2,i_3)$'s in
$L^{*}_2$ with $i_3=i_3^{'}$, and $n-v_3[i_3^{'}]$ is the number of
entries $(i_3,i_4)$'s in $L_3$ with $i_3=i_3^{'}$. Thus,
\[(n-u^{*}_2[i_3^{'}])+(n-v_3[i_3^{'}])=|\{(i_2,i_3)\in L^{*}_2|
i_3=i_3^{'}\}|+|\{(i_3,i_4)\in L_3|i_3=i_3^{'}\}.\] When $i_3$ ranges over $1,2,\ldots,n$, we obtain that the
total number of comparisons needed to search $Q_1$ is at most $N(Q_1)=|L^{*}_2|+|L_3|$ (see Fig. \ref{fig:3d}).

Similarly, if $a_{n,i_2,i_3,i_4}\in {Q^{*}_1}$ equals to $x$, it must be the case that $(i_2,i_3)\in S_2$ and
$(i_3,i_4)\in S^{*}_3$, it follows that the total number of comparisons needed to search $Q^{*}_1$ is at most
$N(Q^{*}_1)=|S_2|+|S^{*}_3|$.

Using similar arguments, the numbers of comparisons needed for
searching the above eight 3-dimensional arrays are:
\begin{equation*}
\begin{aligned}
N(Q_1) =  |L_3|+|L^{*}_2|, \ \ \ \ \ \ N({Q^{*}_1}) = |S_2|+|S^{*}_3|,\\
N(Q_2) =  |L_4|+|L^{*}_3|, \ \ \ \ \ \ N({Q^{*}_2}) = |S_3|+|S^{*}_4|,\\
N(Q_3) =  |L_1|+|L^{*}_4|, \ \ \ \ \ \ N({Q^{*}_3}) = |S_4|+|S^{*}_1|,\\
N(Q_4) =  |L_2|+|L^{*}_1|, \ \ \ \ \ \ N({Q^{*}_4}) = |S_1|+|S^{*}_2|.
\end{aligned}
\end{equation*}
Therefore, the total number of comparisons needed for searching
these eight arrays is at most
\[\sum_{k=1}^{4} (|S_k|+|L_k|+|S^{*}_k|+|L^{*}_k|)=4\times 2n^2=8n^2.\]

Steps 1 and 2 leave an $(n-2)^4$ array
\[A_{n-2,4}= \{a_{i_1,i_2,i_3,i_4}|i_1,i_2,i_3,i_4=2,\ldots,n-1\}.\]
Hence, we have for $n>2$,
\begin{equation*}
\tau(n,4)  \leq  \tau(n-2,4)+8n^2+8(2n-1)
\end{equation*}
From this recursion we can get (see Equation \ref{equ:general} for
the derivation)
\begin{equation}
\tau(n,4)\leq \frac{4}{3}n^3+O(n^2).
\end{equation}

\section{Searching $d$-Dimensional Arrays}

The algorithm for 4-dimensional arrays can be generalized to higher
dimensions ($d\geq 4$). The main idea is quite similar: the $2d$
``surfaces" ($(d-1)$-dimensional arrays) of $A_{n,d}$ can be
searched using $2d n^{d-2}+O(n^{d-3})$ comparisons. We achieve this
in two steps. First, select $2d$ special $(d-2)$-dimensional arrays
and partition each of them into two subsets $S$ and $L$. Second, we
search the $2d$ ``surfaces". The subsets $\{S,L\}$ will help cut
some part of each surface, i.e., reduce the comparison number. In
particular, if we fix $(d-3)$ subscripts, the resulting part is a
smaller matrix (in contrast to an $n\times n$ matrix). An $a\times
b$ matrix can be searched using at most $a+b$ comparisons (Lemma
\ref{searchmatrix}), adding them up for all the $2d$ ``surfaces", we
can get the desired upper bound.

First we describe how to select and partition the
$(d-2)$-dimensional arrays. Define $M_1=\{a_{i_1,i_2,\ldots,i_d}\in
A_{n,d}|i_{d-1}=1,i_d=n\}$. Consider the case where $x \not \in
M_1$. For fixed $i_2=i_2^{'},i_3=i_3^{'},\ldots, i_{d-3}=i_{d-3}^{'}$
(where $i_2^{'},i_3^{'},\ldots,i_{d-3}^{'} \in \{1,2,\ldots,n\}$ are constants) we can get
two integer arrays $u[n]$ and $v[n]$ such that
\begin{equation*}
\begin{aligned}
a_{i_1,i_2,\ldots,i_d}|_{i_2=i_2^{'},\ldots, i_{d-3}=i_{d-3}^{'};\
i_{d-1}=1,i_d=n}<x & & \mbox{ for $i_1\leq u[i_{d-2}]$;}
\\
\\
a_{i_1,i_2,\ldots,i_d}|_{i_2=i_2^{'},\ldots, i_{d-3}=i_{d-3}^{'};\
i_{d-1}=1,i_d=n}>x & & \mbox{ for $i_1>u[i_{d-2}]$;}
\\
\\
a_{i_1,i_2,\ldots,i_d}|_{i_2=i_2^{'},\ldots, i_{d-3}=i_{d-3}^{'};\
i_{d-1}=1,i_d=n}<x & & \mbox{ for $i_{d-2}\leq v[i_1]$;}
\\
\\
a_{i_1,i_2,\ldots,i_d}|_{i_2=i_2^{'},\ldots, i_{d-3}=i_{d-3}^{'};\
i_{d-1}=1,i_d=n}>x & & \mbox{ for $i_{d-2}>v[i_1]$;}
\end{aligned}
\end{equation*}

by using the algorithm \textit{Partitioning 2-Dimensional Array}, at
most $2n-1$ comparisons are needed for each fixed $i_2,\ldots
i_{d-3}$. Thus using at most $n^{d-4}  (2n-1)$ comparisons we can
get two integer arrays $u_1$ and $v_1$ of sizes $n^{d-3}$ such
that\\
-- If $i_1\leq u_1[i_2,\ldots,i_{d-2}]$, then
$a_{i_1,i_2,\ldots,i_d}|_{i_{d-1}=1,i_d=n}<x$.\\
Otherwise, $a_{i_1,i_2,\ldots,i_d}|_{i_{d-1}=1,i_d=n}>x$.\\
-- If $i_{d-2}\leq v_1[i_1,\ldots,i_{d-3}]$, then
$a_{i_1,i_2,\ldots,i_d}|_{i_{d-1}=1,i_d=n}<x$.\\
Otherwise, $a_{i_1,i_2,\ldots,i_d}|_{i_{d-1}=1,i_d=n}>x$.

Thus, we can partition $[n]^{d-2}$ into two subsets $S_1$ and $L_1$ such that\\
-- $a_{i_1,i_2,\ldots,i_d}|_{i_{d-1}=1,i_d=n}<x$ for $(i_1,\ldots,i_{d-2})\in S_1$.\\
-- $a_{i_1,i_2,\ldots,i_d}|_{i_{d-1}=1,i_d=n}>x$ for
$(i_1,\ldots,i_{d-2})\in L_1$.

Obviously, we have \\
-- $(i_1,\ldots,i_{d-2})\in S_1$ if and only if $i_1\leq
u_1[i_2,\ldots,i_{d-2}]$ (also $i_{d-2}\leq v_1[i_1,\ldots,i_{d-3}]$).\\
-- $(i_1,\ldots,i_{d-2})\in L_1$ if and only if
$i_1>u_1[i_2,\ldots,i_{d-2}]$ (also $i_{d-2}>
v_1[i_1,\ldots,i_{d-3}]$).

Next we describe the algorithm for searching $d$-dimensional arrays
$A_{n,d}$, for $d\geq 4$. Without loss of generality, we consider
the case where $x\not \in A_{n,d}$.

\textit{Step 1.} \ Partition each of the following $2d$ \
$(d-2)$-dimensional arrays into two subsets.
\begin{equation*}
\begin{aligned}
M_k & = & \{a_{i_1,i_2,\ldots,i_d}|i_{k-2}=1,i_{k-1}=n\}:\ \ \ S_k, L_k
\\[1.5mm]
M^{*}_k & = & \{a_{i_1,i_2,\ldots,i_d}|i_{k-2}=n,i_{k-1}=1\}:\ \ \
S^{*}_k, L^{*}_k
\end{aligned}
\end{equation*}

$k=1,2,\ldots,d$ (here $i_{k-2}$ means $i_{(k-2)\ mod\ d}$, and $
i_{k-1}$ means $i_{(k-1)\ mod\ d}$).

We get $2d$ pairs of mutually complementary subsets $S_k$, $L_k$ and
$S^{*}_k$, $L^{*}_k$ with the following properties:

\begin{eqnarray*}
a_{i_1,\ldots,i_d}|_{i_{k-2}=1,i_{k-1}=n}& &<x<a_{j_1,\ldots,j_d}|_{j_{k-2}=1,j_{k-1} = n}\\
\\
 & & \mbox{ for $(i_k,\ldots,i_{k+d-3})\in S_k$ and
$(j_k,\ldots,j_{k+d-3})\in L_k$};
\\
\\
a_{i_1,\ldots,i_d}|_{i_{k-2}=n,i_{k-1}=1}& &<x<a_{j_1,\ldots,j_d}|_{j_{k-2}=n,j_{k-1} = 1}\\
\\
 & & \mbox{ for $(i_k,\ldots,i_{k+d-3})\in S^{*}_k$ and
$(j_k,\ldots,j_{k+d-3})\in L^{*}_k$}
\end{eqnarray*}
$k=1,2,\ldots,d$ (here $i_{k+d-3}$ means $i_{(k+d-3)\ mod\ d}$
etc.).

For the pair $S_k$ and $L_k$, we have two $(d-3)$-dimensional arrays
$u_k$ and $v_k$ such that, if $i_k\leq
u_k[i_{k+1},\ldots,i_{k+d-3}]$ then $(i_{k},\ldots,i_{k+d-3})\in
S_k$ else $(i_{k},\ldots,i_{k+d-3})\in L_k$; if $i_{k+d-3}\leq
v_k[i_{k},\ldots,i_{k+d-4}]$ then $(i_{k},\ldots,i_{k+d-3})\in S_k$
else $(i_{k},\ldots,i_{k+d-3})\in L_k$, $k=1,2,\ldots,d$. Similarly,
we have $u^{*}_k$ and $v^{*}_k$ for the pair $S^{*}_k$ and
$L^{*}_k$, for $k=1,2,\ldots,d$.

In this step, we obtain $4d$\ \ $(d-3)$-dimensional arrays $u_k$, $v_k$, $u^{*}_k$, $v^{*}_k$ ($k=1,2,\ldots,d$),
using at most $2d  n^{d-4}  (2n-1)$ comparisons.

\textit{Step 2.} \ Search the following $2d$ \ $(d-1)$-dimensional
surfaces of $A_{n,d}$, which are defined by fixing one of the
subscripts to either 1 or $n$.
\begin{equation*}
Q_k = \{a_{i_1,\ldots,i_d}|i_k=1\}\ \ \ \ \ \ \ \ \ \ \ \ Q^{*}_k =
\{a_{i_1,\ldots,i_d}|i_k=n\}
\end{equation*}
$k=1,2,\ldots,d$.

By symmetry, we only need to consider searching
$Q_1=\{a_{1,i_2,\ldots,i_d}\}$.

If $a_{1,\ldots,i_d}\in Q_1$ equals to $x$, we have
$(i_3,\ldots,i_d)\in L_3$ and $(i_2,\ldots,i_{d-1})\in L^{*}_2$. For
fixed $i_3=i^{'}_3,\ldots,i_{d-1}=i^{'}_{d-1}$ (where
$i^{'}_3$,\ldots,$i^{'}_{d-1}\in \{1,2,\ldots,n\}$), there exist two
integers $u=u^{*}_2[i^{'}_3,\ldots,i^{'}_{d-1}]$ and
$v=v_3[i^{'}_3,\ldots,i^{'}_{d-1}]$ such that only when $u<i_2\leq
n$, $(i_2,i^{'}_3,\ldots,i^{'}_{d-1})\in L^{*}_2$; only when
$v<i_d\leq n$, $(i^{'}_3,\ldots,i^{'}_{d-1},i_d)\in L_3$. Thus for
fixed $i_3=i^{'}_3,\ldots,i_{d-1}=i^{'}_{d-1}$, only when $u<i_2\leq
n$ and $v<i_d\leq n$, $a_{1,i_2,i^{'}_3,\ldots,i^{'}_{d-1},i_d}$
possibly equal to $x$. Searching this $(n-u)\times (n-v)$ matrix
needs at most $(n-u)+(n-v)$ comparisons. Notice that $n-u$ is the
number of elements $(i_2,\ldots,i_{d-1})$'s in $L^{*}_2$ with
$i_3=i^{'}_3,\ldots,i_{d-1}=i^{'}_{d-1}$, and $n-v$ is the number of
elements $(i_3,\ldots,i_d)$'s in $L_3$ with
$i_3=i^{'}_3,\ldots,i_{d-1}=i^{'}_{d-1}$. Thus
$(n-u)+(n-v)=|\{(i_2,\ldots,i_{d-1})\in L^{*}_2 |
i_3=i^{'}_3,\ldots,i_{d-1}=i^{'}_{d-1}\}|+|\{(i_3,\ldots,i_d)\in L_3
| i_3=i^{'}_3,\ldots,i_{d-1}=i^{'}_{d-1}\}|$. When
$(i_3,\ldots,i_{d-1})$ ranges over all elements in $[n]^{d-3}$, we
obtain that the total number of comparisons needed to search $Q_1$
is at most $N(Q_1)=|L_3|+|L^{*}_2|$.

Similarly for $Q^{*}_1=\{a_{n,i_2,\ldots,i_d}\}$, we have
$N(Q^{*}_1)=|S_2|+|S^{*}_3|$.

Using similar arguments, the numbers of comparisons needed for
searching the above $2d$ surfaces are
\begin{equation*}
N(Q_k)=|L_{k+2}|+|L^{*}_{k+1}|\ \ \ \ \ \ \ \  \ \ \ \ \
N(Q^{*}_k)=|S_{k+1}|+|S^{*}_{k+2}|
\end{equation*}
$k=1,2,\ldots,d$ (here $L_{k+2}$ means $L_{(k+2)\ mod\ d}$ etc.).

Thus, searching these $2d$ \ $(d-1)$-dimensional surfaces needs at
most
\[ \sum_{k=1}^{d}
\{|L_{k+2}|+|L^{*}_{k+1}|+|S_{k+1}|+|S^{*}_{k+2}|\}=d\times 2n^{d-2}=2d n^{d-2} \] comparisons.

Steps 1 and 2 leave an $(n-2)^d$ \  $d$-dimensional array
\[ A_{n-2,d}=\{a_{i_1,\ldots,i_d}|i_1,\ldots,i_d=2,\ldots,n-1\}.\]
Hence the generalized recursion is
\begin{eqnarray*}
\tau(1,d) & = & 1\\
\tau(2,d) & \leq & 2^d\\
\tau(n,d) & \leq & \tau(n-2,d)+2d n^{d-2}+2d n^{d-4}(2n-1)\ \ \ \ \mbox{for $n>2$}
\end{eqnarray*}

From the recursion, for fixed $d$ there exist a constant $C$ and
$\varepsilon\in \{1,2\}$ such that

\begin{eqnarray*}
\tau(n,d) & \leq & \tau(n-2,d)+2d  n^{d-2}+4d  n^{d-3}\\
          & \leq & \tau(n-4,d)+2d  (n^{d-2}+(n-2)^{d-2})+4d  (n^{d-3}+(n-2)^{d-3})\\
          & \leq & \ \ \ \ \ \cdots\cdots\cdots\cdots\cdots\cdots\cdots\cdots\cdots\cdots\cdots\cdots\cdots\cdots\cdots\\
          & \leq & C+2d  (n^{d-2}+(n-2)^{d-2}+\ldots+{\varepsilon}^{d-2})+
                   4d  (n^{d-3}+(n-2)^{d-3}+\ldots+{\varepsilon}^{d-2})\\
          & \leq & C+d  ((n+1)^{d-2}+n^{d-2}+(n-1)^{d-2}+\ldots+1^{d-2})+4d
          n^{d-2}\\
%\end{eqnarray*}
%\begin{eqnarray*}
          & \leq & C+d \times \int_1^{n+2} t^{d-2}\ dt  +4d  n^{d-2}\\
          &  =   & C+\frac{d}{d-1}(n+2)^{d-1}-\frac{d}{d-1}+4d  n^{d-2}\\
          &  =   & \frac{d}{d-1}n^{d-1}+O(n^{d-2})
\end{eqnarray*}
Therefore,
\begin{equation}
\label{equ:general}
\tau(n,d)\leq \frac{d}{d-1}n^{d-1}+O(n^{d-2}) \
\ \ \ \ \ \ \mbox{$d=4,5,\ldots$}
\end{equation}

The following theorem summarizes our main results.
\begin{theo}
For $n\geq 1$ and $d \ge 4$, $\tau(n,d) \le \frac{d}{d-1}n^{d-1}+O(n^{d-2})$.\\
Specially for $d=4$, $\frac{4}{3}n^{3}-\frac{n}{3}\leq \tau(n,4)
\leq \frac{4}{3}n^{3}+O(n^{2})$.
\end{theo}

\section{Discussions}

In this paper we give an algorithm searching monotone
$d$-dimensional ($d\geq 4$) arrays $A_{n,d}$, which requires at most
$\frac{d}{d-1}n^{d-1}+O(n^{d-2})$ comparisons.  For $d=4$, it is
optimal up to the lower order terms.

For $d=5$, let $D(n,5)=\{(i_1,i_2,i_3,i_4,i_5) \in [n]^5
|i_1+i_2+i_3+i_4+i_5=\lfloor
\frac{5}{2}(n+1)\rfloor\}\cup\{(i_1,i_2,i_3,i_4,i_5)\in
[n]^5|i_1+i_2+i_3+i_4+i_5=\lfloor \frac{5}{2}(n+1)\rfloor+1\}$, then
$|D(n,5)|$ can be calculated to be $\frac{115}{96}n^4+O(n^3)$, which
is the best lower bound on $\tau(n,5)$ currently known. However,
applying the techniques in this paper, a $\frac{115}{96}n^4+O(n^3)$
search algorithm for $A_{n,5}$ hasn't been found (our algorithm
requires $\frac{5}{4}n^4+O(n^3)$ comparisons in the worst case). So
it may be interesting to tighten the bounds for $d>4$.

\end{document}